\documentclass[12pt,a4paper]{article}
\usepackage[T1]{fontenc}
\usepackage[utf8]{inputenc}
\usepackage{t1enc}	
\usepackage{amsmath}
\usepackage{physics}
\usepackage{authblk}
\usepackage[top=2cm, bottom=2cm, left=2.5cm, right=2cm]{geometry}	
\usepackage{color}
\usepackage[normalem]{ulem}
\usepackage{graphicx}
\usepackage{caption}
\usepackage{subcaption}
\usepackage{float}
\usepackage{wrapfig}
\graphicspath{ {/Home/Documents/Dolgozat/MSc/article/} }
\usepackage{siunitx}
\usepackage{textcomp}
\usepackage{fancyhdr}
\usepackage{cite}
\title{Unitarity in Multi-Higgs Production}
\author[a, b]{A. Curko\footnote{curko.arpad@wigner.mta.hu}}
\author[c]{G. Cynolter\footnote{cyn@general.elte.hu}}
\affil[a]{Institute for Theoretical Physics, Eötvös Loránd University,
Budapest, 1117 Pázmány Péter sétány 1/A, Hungary}
\affil[b]{Department of Quantum Optics and Quantum Information,
Institute for Solid State Physics and Optics,
Wigner Research Centre for Physics of the Hungarian Academy of Sciences, 1121 Konkoly-Thege Miklós út 29-33., Hungary}
\affil[c]{MTA-ELTE Theoretical Physics Research Group, Eötvös Loránd University,
Budapest, 1117 Pázmány Péter sétány 1/A, Hungary}

\date{}                     
\begin{document}
\maketitle
\begin{abstract}
It has long been known that perturbative calculations in scalar multi-particle production could break down since fast growing amplitudes appear. 
A recent calculation in the regime $ \lambda n \gg 1 $, where $ n $ is the multiplicity and $ \lambda $ is the self-coupling, gives an amplitude which grows exponentially with the energy, resulting in a divergent propagator and leading to the violation of perturbative unitarity. In this paper the transition rate is calculated from the solution of the self-consistent Schwinger-Dyson equation (SDE) in spectral representation. We get an amplitude growing quadratically with the energy which leads to an asymptotically decreasing propagator contrary to previous results. Hence, unitarity is not violated, as expected in the Standard Model (SM).
\end{abstract}
\section{Introduction}
Perturbative unitarity is a very useful concept in the SM of electroweak interactions. It shows that the Higgs boson is necessary in the SM beside the weak gauge bosons and also provided a no-go theorem, that the LHC should have found the Higgs boson or sign of new physics below 1 TeV. The Higgs boson was discovered with a mass 125 GeV meaning that the whole SM is perturbative from the electroweak scale up to the Planck scale. The SM describes successfully the high energy experiments but odd behaviours within the SM could point towards new physics. The metastability of the SM vacuum at $10^{10}$-$10^{12}$ GeV is one scale where new physics could appear \cite{Degrassi}.

Recently in \cite{Khoze}, \cite{Khoze2} it was found that in the near threshold production of multi self-interacting scalars (Higgses) there is an unlimited growth of the $ h^* \rightarrow n \times h $ amplitude questioning unitarity. Khoze et al. proposed the so called 'Higgspersion' mechanism \cite{Khoze} where the highly excited Higgs ($ E \gg m $) can only appear as an internal, virtual particle, thus beside the ever growing amplitude this scalar propagator has also a contribution to the cross section. Unitarity is not violated if the propagator can be re-summed, in which case it becomes asymptotically well-behaved, so that it can supress the  high-energy contribution of the amplitude.
However, the reliability of the results of \cite{Khoze} were critized in \cite{Belyaev}, \cite{Monin}, where it was argued that the propagator diverges and cannot be  resummed in the given form because of the exponential growth of the amplitude and the self-energy. 
Using the calculated transition rate outside the region of validity (from the point of view of the steepest-descent method applied in \cite{Khoze}, \cite{Khoze2}) may be one of the reasons of the exponentially growing result. To resolve the issue, we have studied the transition rate with another non-strictly perturbative calculation solving the SDE in the dispersion represantion of the propagator and the self-energy, assuming that this perturbatively re-summed solution does not lead to the violation of unitarity.  We have found that instead of an exponential there is only an $E^2$ dependence in the amplitude. This way we get an asymptotically decreasing propagator and perturbative unitarity is recovered.

The rest of paper is organized as follows. In Sec. 2 we review the multi-Higgs production, then in Sec. 3 we derive the SDE for the spontaneously broken $\Phi^4$ model and we solve it numerically in spectral representation. We discuss unitarity of the multi-Higgs production analyzing the high-energy behaviour of the Higgs propagator and of the amplitude in Sec.4  and we conclude in Sec. 5.
\section{Multi-Higgs production}
Our aim is to investigate the multi-particle production rate of self-interacting scalars using a different method (solving the SDE) than in the previous calculations \cite{Khoze}, \cite{Khoze2}, \cite{Khoze3, Khoze4, Brown, Libanov, TSon, Jaeckel}. The main task is to determine the transition rate of a highly virtual Higgs boson ($ E \gg m $) which decays into $ n $ Higgses. \\
We are going to consider a simplified model of the SM Higgs boson with a single real scalar field $ h(x) $ which has a nonvanishing vacuum expectation value (VEV):
\begin{equation}\label{eq1a} 
\mathcal{L} = \frac{1}{2}\partial^\mu h(x) \partial_\mu h(x) - \frac{\lambda}{4} \left(h(x)^2 - v^2 \right)^2.
\end{equation}
The physical scalar field $ \varphi(x) = h(x) - v $ describes the massive Higgs boson:
\begin{equation}\label{eq2a} 
\mathcal{L} = \frac{1}{2}\partial^\mu \varphi(x) \partial_\mu \varphi(x) - \frac{m^2}{2} \varphi^2(x) - \frac{\kappa}{3} \varphi^3(x) - \frac{\lambda}{4} \varphi^4(x),
\end{equation}
where $ m = \sqrt{2 \lambda} v $ and $ \kappa = 3 \lambda v $. The relevant quantity that characterizes the $ 1 \rightarrow n $ process is the $ R_n(p^2) $ transition rate:
\begin{equation}\label{eq3a}
\int d \Pi_n |\mathcal{M}(1 \rightarrow n)|^2 = R_n(p^2),
\end{equation}
where the $ |\mathcal{M}|^2 $ scattering amplitude squared is integrated over the $ n $-particle phase space $ d \Pi_n $. Recently in \cite{Khoze}, \cite{Khoze2} the transition rate was calculated in the $ \lambda n \gg 1 $ limit using the steepest-descent method, following the approach of \cite{TSon}:
\begin{equation}\label{eq4a} 
R_n(p^2) \propto \exp \left[ n \left( \log \left( \frac{\lambda n}{4} \right) - 1 + \frac{3}{2} \left( \log \left( \frac{\varepsilon}{3 \pi} \right) + 1 \right) - \frac{25}{12}\varepsilon + 0.85 \sqrt{\lambda n}  \right) \right],
\end{equation}
where $ \varepsilon  = (E - mn)/mn $ is the average kinetic energy of the final-state Higgs particles. It is worth mentioning that the above expression within the framework of the steepest descent method is valid only in the $ \lambda \rightarrow 0 $, $ \lambda n = \text{fixed} \gg 1 $, $ \varepsilon = \text{fixed} \ll 1 $ limit. The first terms in (\ref{eq4a}) correspond to the tree-level multiparticle rate, while the last term, which goes with $ \exp (n\sqrt{\lambda n}) $ is the quantum correction and comes from the thin-wall approximation \cite{Khoze2} at the kinematical threshold ($ \varepsilon = 0 $). As one can see from (\ref{eq4a}), the peak rate grows exponentially with the $ E $ energy of the initial Higgs boson, keeping constant the $ \lambda $ coupling. However, in this case none of the conditions $ \lambda n = \text{fixed} \gg 1 $, $ \varepsilon = \text{fixed} \ll 1 $ are satisfied. \\
This exponential growth leads to the question, why does it seem that perturbative unitarity is violated? This is the main issue, which we are going to discuss in the rest of this paper.

\section{Transition rate from the Schwinger-Dyson equations}

Using the path-integral formalism, starting from the identity:
\begin{equation}\label{eq1} 
\int \mathcal{D} \varphi \frac{\delta\theta[\varphi]}{\delta\varphi(x)} = 0 \text{, where } \theta[\varphi] = \exp \left\lbrace  i \int d^4 x \left[  \mathcal{L}(\varphi(x)) + J(x) \varphi(x) \right] \right\rbrace,
\end{equation}
we can get an infinite tower of coupled equations, namely the Schwinger-Dyson equations \cite{Itzykson}.

In the case of the spontaneously symmetry breaking (SSB) $ \varphi^4 $ theory (see equation (\ref{eq2a})) one can arrive at the following self-consistent equation \cite{Sauli}:
\begin{equation}\label{eq2}
\begin{split}
& \Pi(p^2) = \kappa \int \frac{d^4 k_1}{(2 \pi)^4} G_2^c(-k_1) G_2^c(k_1+p) \Gamma_3(k_1, p)  + \\
&+ \lambda \int \frac{d^4 k_1}{(2 \pi)^4} \int \frac{d^4 k_2}{(2 \pi)^4} G_2^c(k_1) G_2^c(-k_1-k_2) G_2^c(k_2+p) \Gamma_4(-k_1, k_1+k_2, p) + \\
&+ 3 \lambda \int \frac{d^4 k_1}{(2 \pi)^4} \int \frac{d^4 k_2}{(2 \pi)^4} G_2^c(-k_1) G_2^c(-k_2) G_2^c(k_1+k_2+p) \Gamma_3(k_1 + p, k_2) G_2^c(k_1+p) \Gamma_3(k_1, p),
\end{split}
\end{equation}
where $ \Pi(p^2) $ is the self-energy, $  G_2^c(k) $ is the connected two-point function, $ \Gamma_n(k_1, k_2, \ldots k_{n-1}) $ is the n-point vertex function, while the constant tadpole term is absorbed by the mass definition. The terms appearing in the SDE can be represented graphically:
\begin{figure}[H]
\centering
\includegraphics[width=0.4\textwidth]{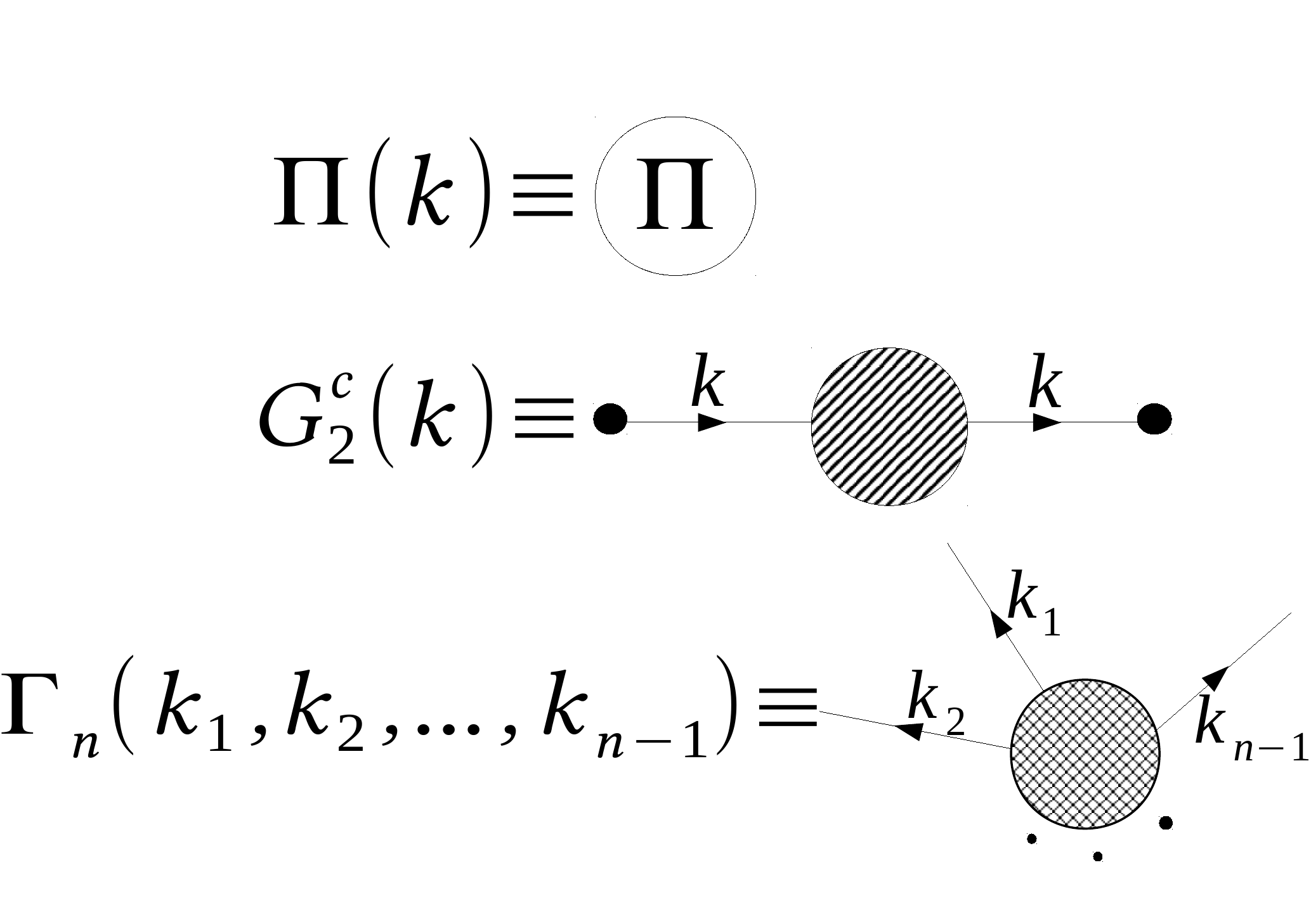}
\caption{Graphical representation of $ \Pi(k) $, $ G_2^c(k) $ and $ \Gamma_n(k_1, k_2, \ldots, k_n) $}
\label{fig1a}
\end{figure}
Using these notations one can write the graphical repesentation of the SDE (\ref{eq2}):
\begin{figure}[H]
\centering
\includegraphics[width=0.6\textwidth]{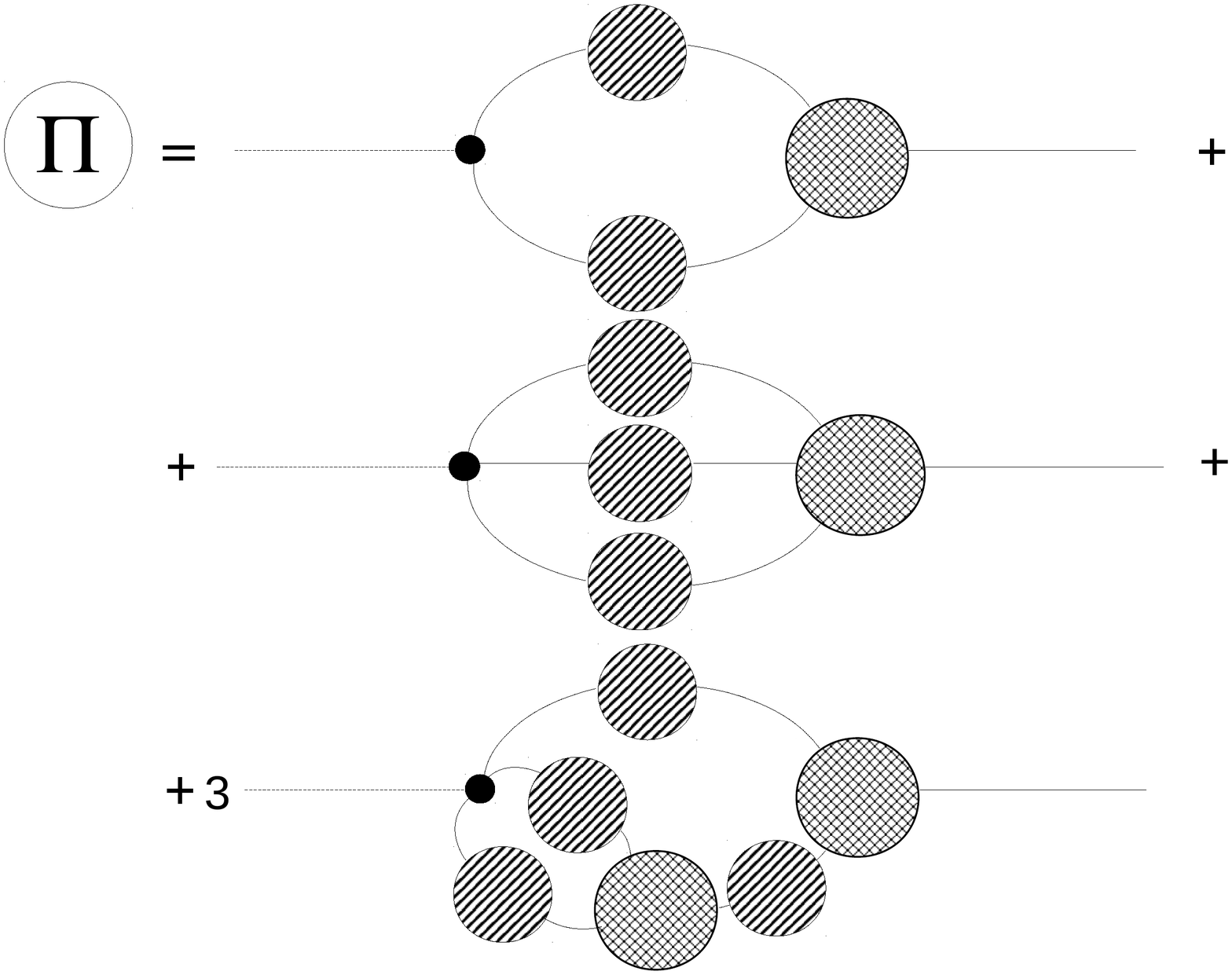}
\caption{The SDE for the $ \varphi^4 $ SSB theory}
\label{fig2a}
\end{figure}
We could write another set of equations for the vertex functions, but we are going to consider only the first terms in further calculations, the other terms would give higher order corrections, which we are going to neglect because of the small value of the  coupling constant $ \lambda $\footnote{$ \lambda = 0.125 $ for the SM Higgs.}:
\begin{equation}\label{eq3} 
\begin{split}
\Gamma_3(k_1, k_2) &= -i 2 \kappa + \ldots , \\
\Gamma_4(k_1, k_2, k_3) &= -i 6 \lambda + \ldots \text{ .}
\end{split}
\end{equation}
Finally one obtains:
\begin{equation}\label{eq4} 
\begin{split}
\Pi(p^2) &=  - i 2 \kappa^2 \int \frac{d^4 k_1}{(2 \pi)^4} G_2^c(-k_1) G_2^c(k_1+p)  - \\
&- i 6 \lambda^2 \int \frac{d^4 k_1}{(2 \pi)^4} \int \frac{d^4 k_2}{(2 \pi)^4} G_2^c(k_1) G_2^c(-k_1-k_2) G_2^c(k_2+p) - \\
&- 12 \lambda \kappa^2 \int \frac{d^4 k_1}{(2 \pi)^4} \int \frac{d^4 k_2}{(2 \pi)^4} G_2^c(-k_1) G_2^c(-k_2) G_2^c(k_1+k_2+p) G_2^c(k_1+p), 
\end{split}
\end{equation}
where the first term corresponds to the so-called bubble diagram, while the second contribution corresponds to the setting sun diagram. Our approach is to solve this SDE, following the calculations of Sauli \cite{Sauli}, \cite{Sauli2}, using the spectral decomposition method. The generic spectral decomposition of the scalar two-point function reads:
\begin{equation}\label{eq5} 
G(p^2) = \frac{i Z}{p^2 - m^2 + i\epsilon} + \int_{\omega_{th}}^{\infty} d \omega \frac{i \sigma(\omega)}{p^2 - \omega + i\epsilon},
\end{equation}
where we assumed that there is a pole term at the physical mass $ m $ ($ Z $ is the wave function renormalization factor) and a regular, continuum term determined by the spectral function $  \sigma(\omega) $. At the beginning let's consider only the first two terms in (\ref{eq4}) and later we will give an order-of-magnitude estimate for the third diagram. In 3+1 dimensions the bubble contribution has a logarithmic divergence, while the setting-sun contribution diverges quadratically. In order to get a finite result we have to make two subtractions by introducing the renormalized self-energy $ \Pi_R(p^2) $, we use the Non-minimal Momentum Subtraction (NMS) renormalization scheme: 
\begin{equation}\label{eq6} 
\Pi_R(p^2) = \Pi(p^2) - \Pi(m^2) - \frac{d \Pi(p^2)}{d p^2}|_{p^2=m^2}(p^2-m^2).
\end{equation} 
Choosing this double subtraction procedure the dispersion relation formula for the renormalized self-energy reads:
\begin{equation}\label{eq7} 
\Pi_R(p^2) = \int_{\omega_{th}}^{\infty} d\omega \frac{\rho(\omega)}{p^2 - \omega + i\epsilon} \left( \frac{p^2-m^2}{\omega - m^2}  \right)^2 = \int_{\omega_{th}}^{\infty} d\omega \frac{\tilde{\rho}(\omega, p^2; m^2)}{p^2 - \omega + i\epsilon}.
\end{equation}
this equation defines the spectral function of the self-energy $ \rho(\omega) = - \text{Im}\Pi_R(p^2)/ \pi $ and also the double substracted spectral function $ \tilde{\rho}(\omega, p^2; m^2) $. Neglecting the bound states we set the threshold of the spectrum variable at $ \omega_{th} = 4 m^2 $ (the bubble diagram contains two propagators). So finally from the SDE using the spectral representations (\ref{eq5}) and (\ref{eq7}) we get a relation between the two spectral functions $ \sigma(\omega) $ and $ \rho(\omega) $:
\begin{equation}\label{eq8} 
\begin{split}
\rho(\omega) &= \frac{2 \kappa^2}{(4\pi)^2} \left[Z^2 X(m^2, m^2, \omega) + 2 Z \int_{\omega_{th}}^{\infty} d\alpha_1 X(\alpha_1, m^2, \omega) \sigma(\alpha_1) + \right.  \\
&+\left.  \int_{\omega_{th}}^{\infty} d\alpha_1 \int_{\omega_{th}}^{\infty} d\alpha_2 X(\alpha_1, \alpha_2, \omega) \sigma(\alpha_1) \sigma(\alpha_2)\right] + \\
&+ \frac{6\lambda^2}{(4\pi)^4} \left[ Z^3 Y(m^2, m^2, m^2, \omega) + 3 Z^2 \int_{\omega_{th}}^{\infty} d\alpha_1  Y(\alpha_1, m^2, m^2, \omega)\sigma(\alpha_1) + \right. \\
& + 3 Z \int_{\omega_{th}}^{\infty} d\alpha_1 \int_{\omega_{th}}^{\infty} d\alpha_2  Y(\alpha_1, \alpha_2, m^2, \omega) \sigma(\alpha_1) \sigma(\alpha_2) + \\
& \left. + \int_{\omega_{th}}^{\infty} d\alpha_1 \int_{\omega_{th}}^{\infty} d\alpha_2 \int_{\omega_{th}}^{\infty} d\alpha_3 Y(\alpha_1, \alpha_2, \alpha_3, \omega)  \sigma(\alpha_1) \sigma(\alpha_2) \sigma(\alpha_3) \right],
\end{split}
\end{equation}
where $ X(\alpha_1, \alpha_2, \omega) $ and $ Y(\alpha_1, \alpha_2, \alpha_3, \omega) $ are purely kinematical functions\footnote{$X(\alpha_1, \alpha_2, \omega)  = \frac{\sqrt{\lambda(\alpha_1, \alpha_2, \omega)}}{\omega} \Theta (\omega - (\sqrt{\alpha_1} + \sqrt{\alpha_2})^2),$ \\ $ Y(\alpha_1, \alpha_2, \alpha_3, \omega) = \int_{\left( \sqrt{\alpha_1} + \sqrt{\alpha_2} \right)^2}^{\left( \sqrt{\omega} - \sqrt{\alpha_3} \right)^2 } ds \frac{\sqrt{\lambda(s, \alpha_3, \omega)} \sqrt{\lambda(s, \alpha_1, \alpha_2)}}{s\omega}\Theta \left( \left( \omega - \sqrt{\alpha_1} + \sqrt{\alpha_2}  + \sqrt{\alpha_2} \right)^2 \right),  $ \\ where the $ \lambda(x, y, z) = x^2 + y^2 + z^2 - 2xy - 2xz - 2yz $ is the Källén function and the $ \Theta(x) $ is the Heaviside step function.}. One can easily show that the third diagram of the equation (\ref{eq4}) would give a further term to the SDE (\ref{eq8}) with a prefactor of $ \frac{12 \lambda \kappa^2}{(4\pi)^4} $, which is smaller by three orders of magnitude than the factor of the bubble diagram. So, as we have already mentioned, we are going to investigate the simplified equation (\ref{eq8}), which contains the bubble and the setting sun diagrams. To solve the above equation we need another relation \cite{Sauli}, it can be derived from the trivial identity $ G(p^2)G^{-1}(p^2) = 1 $ using $ G^{-1}(p^2) = p^2 - m^2 -\Pi_R(p^2) $ and the spectral representations (\ref{eq5}), (\ref{eq7}):
\begin{equation}\label{eq9} 
\sigma (p^2) = \frac{Z \rho(p^2)}{(p^2-m^2)^2} + \frac{1}{p^2-m^2}\text{P} \cdot \int_{\omega_{th}}^{\infty} d\omega \frac{\sigma(p^2)\tilde{\rho}(\omega, p^2; m^2) + \rho(p^2)\sigma(\omega)}{p^2 - \omega}
\end{equation}
where $ \text{P} \cdot $ denotes the Cauchy principal value integral. Beside the two spectral functions there is another unkown variable, namely the wave function renormalization factor $ Z $. This can be easily found from the sum rule: $ Z + \int_{\omega_{th}}^{\infty} d\omega \sigma(\omega) = 1 $, which is another consequence of the above-mentioned trivial identity. Though in this choice of subtraction (see equation (\ref{eq6})) the $ Z $ equals $ 1 $, we have to verify whether $ \int_{\omega_{th}}^{\infty} d\omega \sigma(\omega) \ll 1 $ in order to satisfy approximately the sum rule. 
The complete solution, where the subtraction scheme is modified in such a way that the $ Z $ factor is not set to $ 1 $ and it is calculated iteratively from the sum rule, is in accordance with the result of the NMS scheme if the contribution of the contiunuum term is much smaller than that of the pole term. This is the case here, as we will see later the contribution of the continuum term has a value of $ O(10^{-3}) $ thanks to the small coupling $ \lambda $.  

Our main goal is to calculate the transition rate of the multi-Higgs production from the SDE. We can achieve this if we relate the  spectral function $ \rho(\omega) $ to the  transition rate summed over the $ n $ particle number $ \sum_n R_n(p^2) $ via the LSZ reduction formula \cite{Peskin}:
\begin{equation}\label{eq10} 
-Z \Pi(p^2) = \mathcal{M}(p \rightarrow p)
\end{equation}
and via the optical theorem \cite{Peskin}:
\begin{equation}\label{eq11} 
2 \text{Im}\mathcal{M}(p \rightarrow p) = \sum_n R_n(p^2).
\end{equation} 
So if we calculate the spectral function of the renormalized self-energy we can obtain the transition rate:
\begin{equation}\label{eq12} 
\sum_n R_n(p^2) = 2 \pi Z \rho(p^2).
\end{equation}

\subsection{Numerical solution}
Before discussing the numerical solution of the SDE, let's investigate the high energy behaviour of the spectral function $ \rho(\omega) $ by dimensional analysis. We know that the spectral function of the self-energy must have the following form according to the equation (\ref{eq4}):
\begin{equation}\label{eq13} 
\rho(\omega) = m^2 \lambda f_1(\omega, m^2) + \lambda^2 f_2(\omega, m^2) +  m^2 \lambda^2 f_3(\omega, m^2),
\end{equation}
where $ f_i(\omega, m^2) $ are spectral functions coming from the particular diagrams. Starting from the fact that the $ \rho(\omega) $ has the dimension $ M^2 $ independently of the chosen subtraction scheme, and the $ \lambda $ coupling is dimensionless, we can easily determine the dimensions of the $ f_i(\omega, m^2) $ functions: $ \left[ f_1(\omega, m^2) \right] = M^0  $,  $ \left[ f_2(\omega, m^2) \right] = M^2  $,  $ \left[ f_3(\omega, m^2) \right] = M^0  $. In the high energy limit ($ \omega \gg m^2 $):
\begin{equation}\label{eq14} 
\begin{split}
f_1(\omega, m^2) & \overset{\omega \gg m^2}{\rightarrow} f_1(\omega) \propto \omega^0, \\
f_2(\omega, m^2) & \overset{\omega \gg m^2}{\rightarrow} f_2(\omega) \propto \omega, \\
f_3(\omega, m^2) & \overset{\omega \gg m^2}{\rightarrow} f_3(\omega) \propto \omega^0.
\end{split}
\end{equation}
According to (\ref{eq14}) above high enough energy, no matter what kind of prefactors belong to the diagrams, the setting sun contribution will determine the behaviour of the $ \rho(\omega) $, this will suppress the other two contributions because of the $ \omega $ dependence. Knowing this later we can verify the validity of our numerical calculations. \\
From the unitarity relation (\ref{eq9}) we know that $  \sigma(\omega) > \frac{\rho(\omega)}{(\omega - m^2)^2} $, furthermore at small couplings, as we will see later, $  \sigma(\omega) \approx \frac{\rho(\omega)}{(\omega - m^2)^2} $ is also true. In the $ \int_{\omega_{th}}^{\infty} d\omega \sigma(\omega) $ integral a logarithmic divergence will appear because of the setting sun diagram, but this contradicts the sum rule. It is important to emphasize that this problem is independent of the renormalization/subtraction, because the equation (\ref{eq13}) can be written for spectral functions ($ \rho $, $ f_i $) after arbitrary $ n $ subtractions. To solve this problem, we have to take into account that the $ \varphi^4 $ is a trivial theory: if we introduce a cut-off $ \Lambda $ in the theory and take the $ \Lambda \rightarrow \infty $ limit, the renormalized coupling goes to zero, we will get a free theory without interaction. For that reason if we define this $ \varphi^4 $ theory up to a cut-off $ \Lambda $, the problem of the $ \int d\omega \sigma(\omega) $ integral will be solved. We choose the value of $ \Lambda $ in such a way that it will be well above the energy range which we would like to investigate and to get a  $ \int_{\omega_{th}}^{\Lambda} d\omega \sigma(\omega) \ll 1 $ integral according to the aforementioned sum rule. So in the further calculations we will use the arbitrarily chosen\footnote{We have found that various values of the cutoffs give the same $ \rho $ and $ \sigma $ spectral functions.} $ \Lambda = (800 m)^2 $, which corresponds to the energy $ E_{max} = 100 $ TeV.

\begin{figure}[H]
\centering
\includegraphics[width=0.7\textwidth]{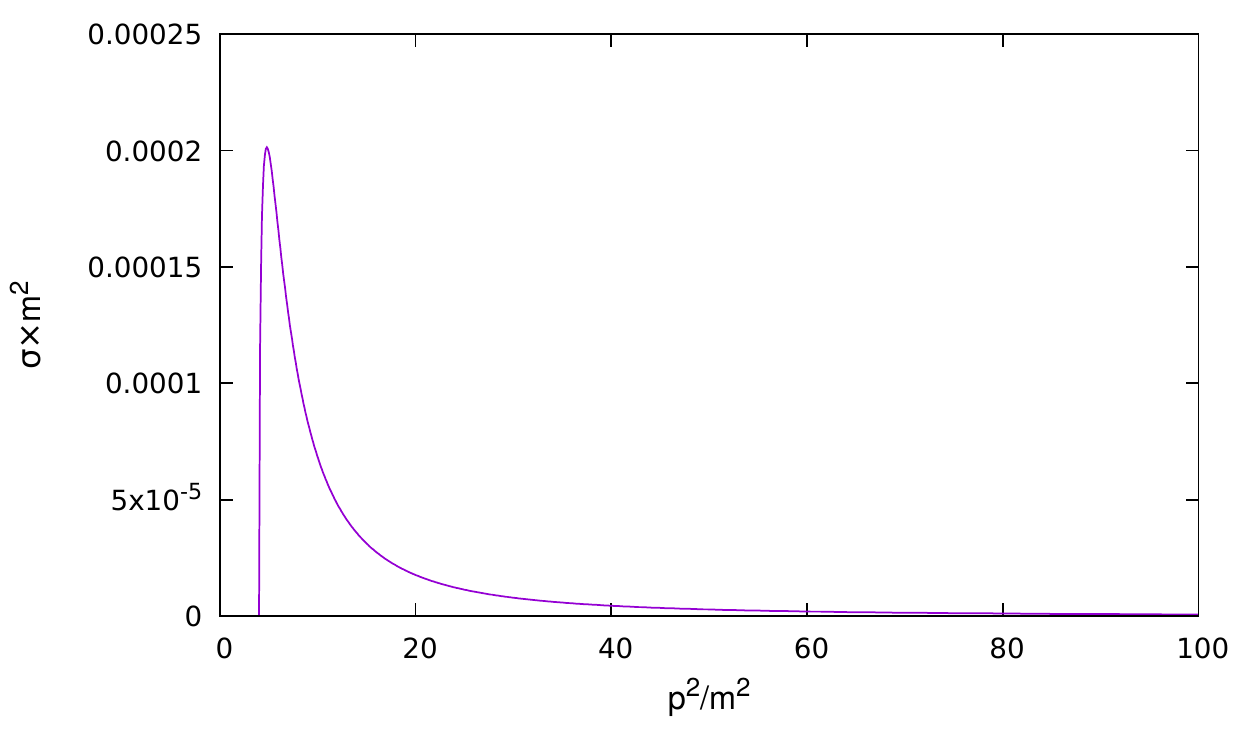}
\caption{The $ \sigma $ spectral function of the propagator; $ \lambda = 0.125 $}
\label{fig1}
\end{figure}

\begin{figure}[H]
\centering
\includegraphics[width=0.7\textwidth]{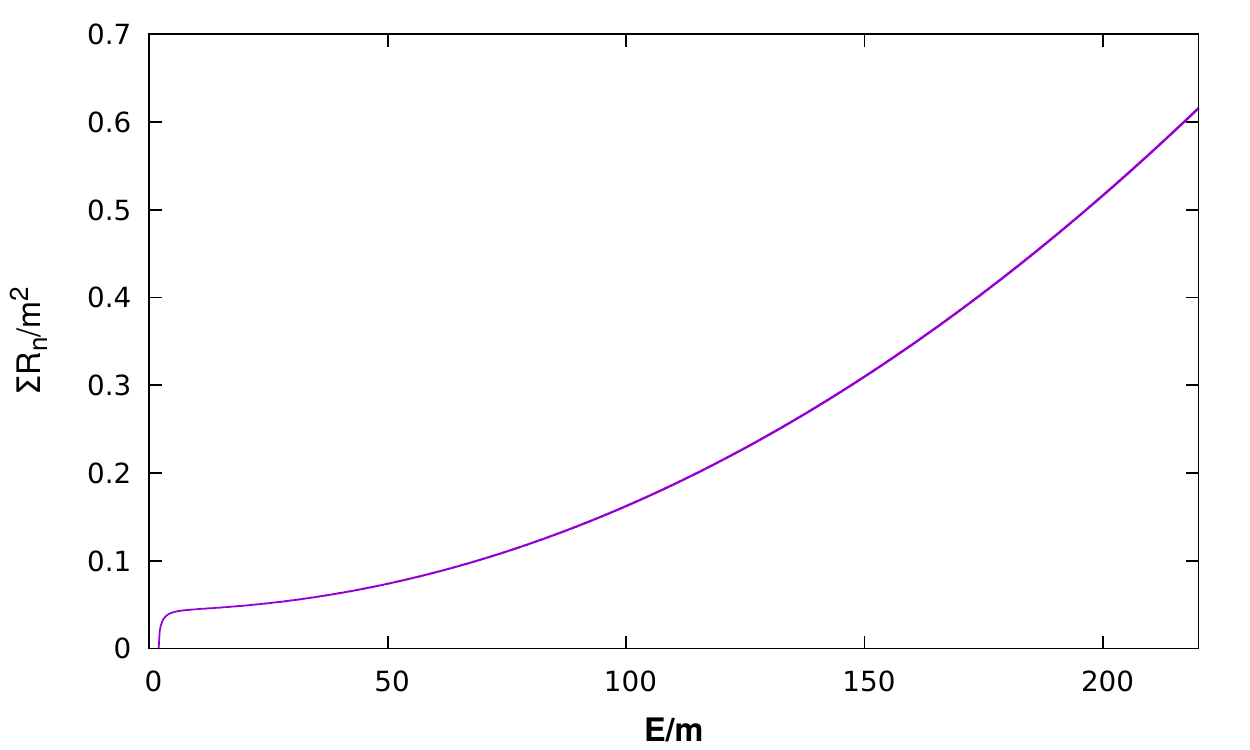}
\caption{The $ \sum_n R_n(E) $ transition rate at high energies; $ \lambda = 0.125 $}
\label{fig2}
\end{figure}

\begin{figure}[H]
\centering
\begin{subfigure}{0.5\textwidth}
\caption{$ \lambda = 0.125 $}
\includegraphics[width=1\textwidth]{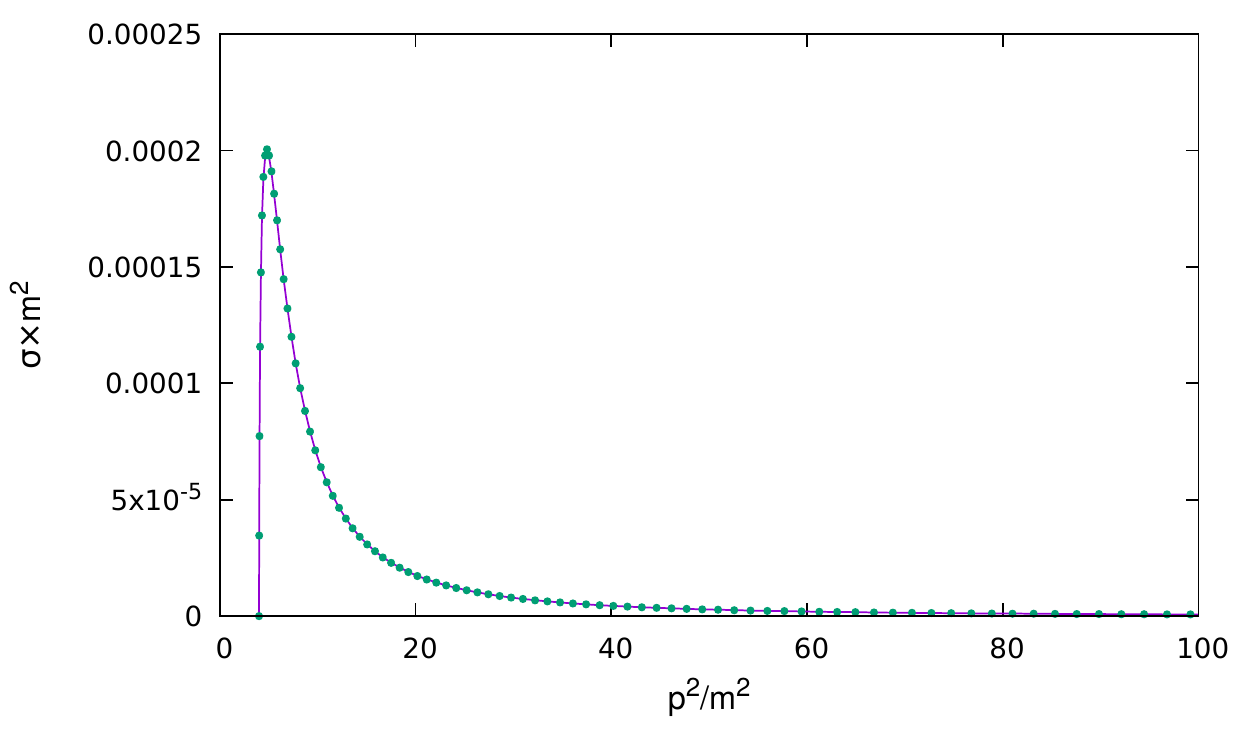}
\label{fig3a} 
\end{subfigure}%
~
\begin{subfigure}{0.5\textwidth}
\caption{$ \lambda = 10 $}
\includegraphics[width=1\textwidth]{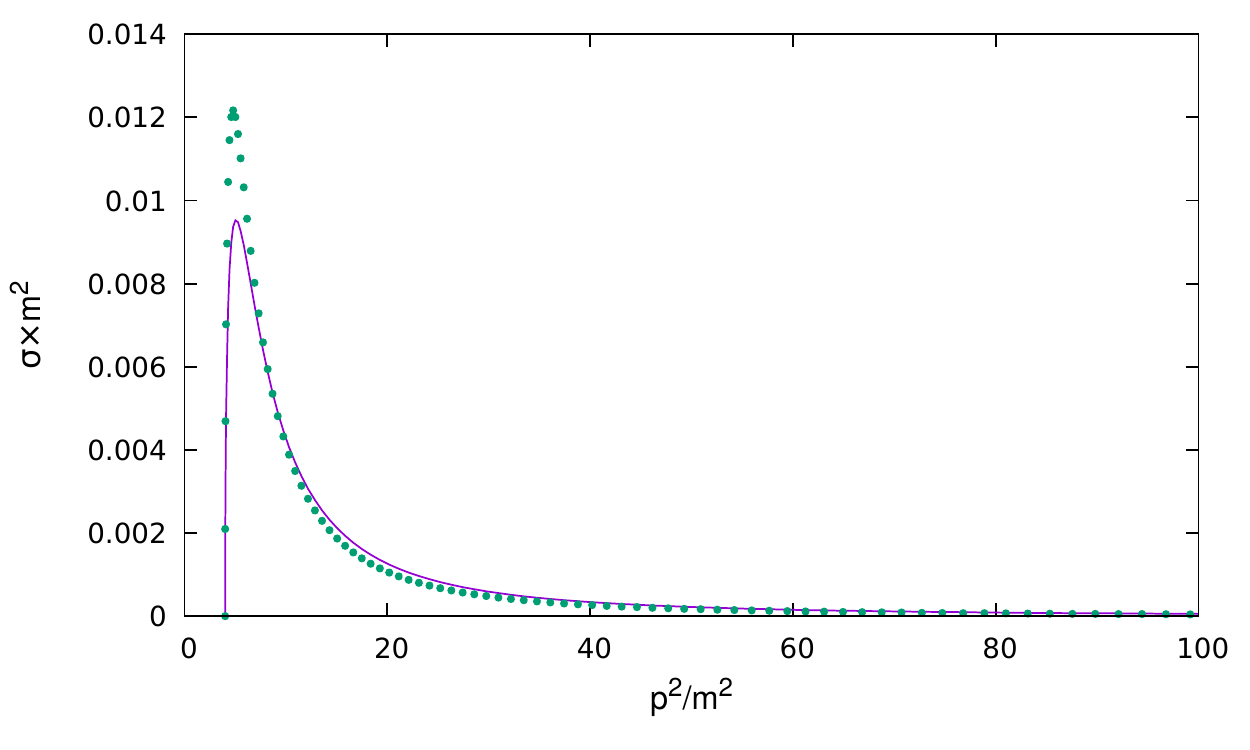}
\label{fig3b}
\end{subfigure}
\caption{The $ \sigma $ spectral function using the whole equations (continuous line) and the simplified equations (points) for small \textbf{(a)} and large \textbf{(b)} values of $ \lambda $}

\end{figure}

The set of equations (\ref{eq8}) and (\ref{eq9}) can be solved iteratively \cite{Marko}. As a first step, we start with a propagator, that contains only the pole term (this means that we set the continuum term $ \sigma(\omega) $ to zero). Then we calculate iteratively the spectral functions $ \rho(\omega) $, $ \sigma(\omega) $ performing the integrals over the $ \left[ \omega_{th}, \Lambda \right]  $ interval. The convergence of the iteration is tested by the integral $ \int_{\omega_{th}}^{\Lambda} d\omega \sigma(\omega) $, this has a value of $ O(10^{-3}) $ in the $ \lambda = 0.125 $ case, which also means that our $ Z \approx 1 $ approximation is valid. For the small coupling $ \lambda = 0.125 $ the spectral function of the propagator, see \textbf{Fig. \ref{fig1}}, starts at the threshold $ p^2 = 4 m^2 $, reaches his maximum at $ \sigma = 0.0002/m^2 $, then goes asymptotically to zero at high energies.
From the numerical calculations we have found that the spectral function $ \rho(\omega) $ grows with $ p^2 $ at $ E \gg m $ energies in accordance with the previous dimensional analysis. Thus the summed transition rate $ \sum_n R_n(E) $ calculated from the spectral function of the renormalized propagator goes with the square of the energy at high energies,  $ E \gg m $  as one can see in \textbf{Fig. \ref{fig2}}. Therefore the transition rate increases continuously up to the cut-off $ \Lambda $, which at a first glance contradicts perturbative unitarity. Before we are going to discuss this problem, let's investigate what happens if we keep only the first terms in the equations (\ref{eq8}) and (\ref{eq9}) (this we are going to call the simplified solution).
At small couplings, see \textbf{Fig. \ref{fig3a}}, the difference between the simplified solution and the result of the whole equations (\ref{eq8}), (\ref{eq9}) lies within the linewidth. This means, as we have mentioned already, that the  $  \sigma(\omega) \approx \frac{\rho(\omega)}{(\omega - m^2)^2} $ approximation is true if $ \lambda \ll 1 $. Nevertheless at large couplings, see \textbf{Fig. \ref{fig3b}}, it is not enough to solve only the simplified equations to get the right result, which still converges but slower than for small couplings.

\section{Unitarity}

\subsection{The Higgs propagator}
We have already used the fact that the propagator can be written as a geometric series of the 1PI (one-particle irreducible) self-energy $ \Pi(p^2) $:
\begin{align}\label{eq15}
\begin{split}
G(p^2) &= \frac{i}{p^2 - m^2} \sum_{n = 0}^{\infty} \left( -i \Pi(p^2) \frac{i}{p^2 - m^2} \right)^n \\
&= \frac{i}{p^2 - m^2 - \Pi(p^2)}.
\end{split}
\end{align}
However, if the series does not converge, the propagator becomes divergent. In \cite{Belyaev} Belyaev et al. have already shown that in the case of the solution (\ref{eq4a}) the power series (\ref{eq15}) is divergent at high energies $ E \gg m $, the self-energy grows exponentially with $ \sqrt{p^2} $. But considering the SDE solution, the spectral function of the self-energy grows only with $ p^2 $, thus the series is convergent at any $ E $ and one can resum the $ \Pi(p^2) $ into the denominator of the propagator. As the self-energy grows, the propagator decreases with the energy, this is also shown by the high-energy behaviour of the spectral function of the propagator $ \sigma(p^2) $ (see \textbf{Fig. \ref{fig1}}). So this solution of the multi-Higgs production problem ensures an asymptotically vanishing spectral density of the propagator (already pointed out in \cite{Monin}).

\subsection{The Higgspersion mechanism}
Our starting point is that the off shell, decaying Higgs boson can only take part in real, physical processes as an intermediate, virtual particle. Let's investigate the following process according to \cite{Khoze}: a high energy, virtual Higgs boson is created in gluon fusion, which then decays into a large number of on shell Higgs bosons. The amplitude of this process reads:
\begin{equation}\label{eq16}
\mathcal{M}_{gg \rightarrow n \times h}(p^2) = \mathcal{M}_{gg \rightarrow h^*}(p^2) \times G(p^2) \times \mathcal{M}_{h^* \rightarrow n \times h}(p^2).
\end{equation}
Using the thin-wall approximation of \cite{Khoze2}, beside the exponentially growing amplitude $ \mathcal{M}_{h^* \rightarrow n \times h}(p^2) $ the $ G(p^2) $ is also divergent at high energies, therefore we get a cross section which violates unitarity. But if the self-energy can be resummed via the power series (\ref{eq15}), one can give an upper bound on the cross section:
\begin{equation}\label{eq17}
\begin{split}
\sigma_{gg \rightarrow n \times h} &\propto | \mathcal{M}_{gg \rightarrow h^*} |^2 \times \frac{1}{\left( p^2 - m^2 - \text{Re} \Pi(p^2) \right)^2 + \left(  \text{Im} \Pi(p^2) \right)^2 } \times R_n(p^2) \\
& < \log^4 \left( \frac{m_t}{\sqrt{p^2}} \right) 4 Z^2 \frac{1}{\sum_n R_n(p^2)},
\end{split}
\end{equation} 
where we have used the equations (\ref{eq10}), (\ref{eq11}) and the fact that the cross section of the $ gg \rightarrow h^* $ process can be written as $ \sigma_{gg \rightarrow h^*} \propto \log^4 \left( \frac{m_t}{\sqrt{p^2}} \right) $ according to \cite{Khoze}\footnote{$ m_t $ is the mass of the top quark.}. As we have seen the solution of the SDE gives a transition rate $ \sum_n R_n(p^2) $ proportional to $ p^2 $, so using (\ref{eq17}) the cross section  $ \sigma_{gg \rightarrow n \times h} $ goes to zero as $ p^2  $ goes to infinity. Finally this real physical process does not lead to the violation of unitarity, even though the amplitude of the $ h^* \rightarrow n \times h $ multi-Higgs production grows continuously with the energy.
\section{Conclusions}
We have investigated the $ 1 \rightarrow n $ multi-Higgs production, bearing in mind the fact that perturbative unitarity should not be violated. We have examined this scattering process by solving numerically the SDE of the $ \varphi^4 $ Higgs model in spectral representation. From the calculations we have found that the summed transition rate $ \sum_n R_n(E) $ goes with the \textit{square} of the energy in the case of a highly virtual initial Higgs boson ($ E \gg m $). However, previously it was argued in \cite{Khoze}, \cite{Khoze2}, in the framework of the thin-wall approximation, that the transition rate of the $ 1 \rightarrow n $ process increases \textit{exponentially} with the energy. In \cite{Khoze}, \cite{Khoze2} it was proposed that the so-called Higgspersion mechanism could restore perturbative unitarity, where the  intermediate particle is a decaying Higgs boson produced via gluon fusion and its propagator can supress the ever growing $1\rightarrow n $ amplitude. But as it was shown in \cite{Belyaev}, the Higgspersion mechanism cannot be applied to the result of \cite{Khoze}, because the propagator becomes divergent at high energies due to the exponential increase of the transition rate. The SDE solution, in turn, gives a propagator which vanishes at high momenta, and thus the cross section of the complete process goes to zero as the energy goes to infinity. So the calculations derived from the SDE show that the perturbative unitarity is not violated in accordance with the renormalizability and unitarity of the Standard Model.

\paragraph{Acknowledgements}\mbox{}\\
We are grateful for Gergő Markó and Zsolt Szép for valuable discussions and useful comments especially on the implementation of numerical calculations.

\end{document}